\documentclass[preprint,12pt]{elsarticle}

\usepackage[export]{adjustbox}
\usepackage{graphicx}
\usepackage{caption}
\usepackage{subcaption}
\usepackage{amsmath}
\usepackage{algorithm}
\usepackage[noend]{algpseudocode}
\usepackage{pgf-pie}
\usepackage{textcomp}
\usepackage[toc,page]{appendix}
\usepackage{float}
\usepackage[colorinlistoftodos]{todonotes}
\usepackage{cleveref}
\usepackage{ulem}
\usepackage{listings}
\usepackage{xcolor}
\newcommand\editcolor{black}

\begin{document}

\begin{frontmatter}

\title{Real-world K-Anonymity Applications: the \textsc{KGen} approach and its evaluation in Fraudulent Transactions}

\author[JADS]{Jessica {De Pascale}}
\author[EIN]{Giuseppe Cascavilla}
\author[EIN]{Damian A. Tamburri}
\author[JADS]{Willem-Jan Van Den Heuvel}

\address[JADS]{Tilburg University - Jheronimus Academy of Data Science, 's-Hertogenbosch, The Netherlands}
\address[EIN]{Eindhoven University of Technology - Jheronimus Academy of Data Science, 's-Hertogenbosch, The Netherlands}

\begin{abstract}
K-Anonymity is a property for the measurement, management, and governance of the data anonymization. Many implementations of k-anonymity have been described in state of the art, but most of them are not able to work with a large number of attributes in a ``Big'' dataset, i.e., a dataset drawn from Big Data. 
To address this significant shortcoming, we introduce and evaluate \textsc{KGen} an approach to K-anonymity featuring Genetic Algorithms. \textsc{KGen} promotes such a meta-heuristic approach since it can solve the problem by finding a pseudo-optimal solution in a reasonable time over a considerable load of input. \textsc{KGen} allows the data manager to guarantee a high anonymity level while preserving the usability and preventing loss of information entropy over the data. Differently from other approaches that provide optimal global solutions catered for small datasets, \textsc{KGen} works properly also over Big datasets while still providing a good-enough solution.
Evaluation results show how our approach can still work efficiently on a real world dataset, provided by Dutch Tax Authority, with 47 attributes (i.e., the columns of the dataset to be anonymized) and over 1.5K+ observations (i.e., the rows of that dataset), as well as on a dataset with 97 attributes and over 3942 observations.
\end{abstract}


\end{frontmatter}

\section{Introduction}\label{sec:introduction}



The amount of data being produced and processed, both online and offline, is exponentially increasing, and so is the costly consumption of resources to carry such processing to fruition. On the one hand, maintaining \emph{data anonymity} is a must-have, especially in sight of the severe sanctions connected to potential violations of the General Data Protection Regulation \cite{ArfeltBD19}. On the other hand, many agencies want or need to exploit such data for commercial purposes or public safety and security, implying that data should be \emph{usable}. 

It is, hence, fundamental to provide fast and reliable techniques to the stakeholders that guarantee the privacy and anonymity of the data and, at the same time, maintain the data's usefulness. 
This paper introduces and evaluates \textsc{KGen}, an approach to state-of-the-art privacy-preserving technologies implemented using a metaheuristic-based approach.


The process starts with a dataset, and, through an anonymization process, it provides a dataset anonymized. At the core of \textsc{KGen} is the most widely known k-anonymity approach to anonymization \cite{samarati2001microdata}. 
\textcolor{\editcolor}{K-anonymity is defined as the condition wherefore, for each record in that dataset, there are at least other k-1 records indistinguishable from it.}

The K-anonymity property is classified as an NP-Hard problem, as proved by Meyerson et al; \cite{meyerson2004complexity}. Aggarwal \cite{aggarwal2005k} shows the problem raised by any K-anonymity algorithms applied with large datasets. \textcolor{\editcolor}{The information loss of a dataset also depends on the size of a dataset. If the size of a dataset increases, the information loss of a dataset increases too, leading to having a useless dataset with a higher level of anonymization.}


Though it is not possible to anonymize a large dataset without loss of information, with \textsc{KGen} we aim to provide an anonymized dataset on the K-Anonymity property. 
In the scope of \textsc{KGen}, K-anonymity needs to be traded-off against the usefulness of data. At the same time, several algorithms address this problem, providing an optimal solution \cite{samarati2001microdata, sweeney1997guaranteeing, sweeney2002achieving, el2009globally, lefevre2005incognito}, all known approaches merely work on a relatively small number of attributes with a reduced level of generalization for each attribute. While the number of attributes that need to be anonymized grows, the higher is the complexity to obtain a \emph{usable} dataset.

To account for the trade-off mentioned above, \textsc{KGen} features an approach based on Genetic Algorithms \cite{goldberg1988genetic} 
providing \textcolor{\editcolor}{a pseudo-optimal} 
solution in a time \textcolor{\editcolor}{useful for practical usage (in the result of this work the maximum time reached is 2 hours for the dataset with 15 attributes).} 
We compared \textsc{KGen} with other approaches from the state-of-the-art in order to validate its results.

The main goal of this work is to provide an approach useful in an industrial context. To this end, we defined the following research question:



\begin{center}
    \textit{\textbf{Main RQ$_1$:} Is the performance of the proposed approach
    useful for stakeholders?}
\end{center}

To answer the main research question, we outlined three subsequent research questions:

\begin{enumerate}
    \item[RQ1] \textit{Does \textsc{KGen} perform when compared to state-of-the-art approaches?} To address this RQ We first compared our approach to existing ones by means of execution time to generate the best-anonymized dataset.
    
    \item[RQ2] \textit{How accurate are \textsc{KGen} solutions compared to state-of-the-art approaches?} To answer this question, we proposed a measure of accuracy to measure how the pseudo-optimal solution is far from the optimal solution.
    
    \item[RQ3] \textit{What is the quality of \textsc{KGen} solution?} We measured the quality of a solution using generalization and suppression metrics defined in the state-of-the-art and discussed in the Sec. \ref{sec:generalization-suppression}.
    
\end{enumerate}

Moreover, to evaluate the applicability in a large context scenario, we outlined a followup main research question:

\begin{center}
    \textit{\textbf{Main RQ$_2$:} To what extent can the case-specific evaluation generalise to much larger datasets?}
\end{center}

Therefore, in order to evaluate \textsc{KGen} in an industrial context, the approach was used a real-world sample dataset provided by the Dutch Tax Authority for fraudulent transactions. The evaluation aims at accounting for \textsc{KGen}'s real-life applicability. 
Moreover, we led a second experimentation, using the ``c2k\_data\_comma.csv'' dataset \cite{cargo2000dataset} to prove the applicability of the approach using a large dataset.
\textcolor{\editcolor}{The experimentation has been done using OLA \cite{el2009globally}, a state-of-the-art approach for the dataset k-anonymization, a brute force approach and a meta-heuristic random approach to evaluate the goodness of \textsc{KGen}.}
The experimentation reveals promising results and shows that \textsc{KGen} is an approach capable of providing \textcolor{\editcolor}{a good-enough solution in less than 5h:05m:40s (the worst case recorded with the ``c2k\_data\_comma.csv'' dataset and 25 quasi-identifiers attributes to anonymize.} 
\textsc{KGen} showed to be able to find results up to 25 attributes to anonymize, under the limited-time set of 15 hours differently from other approaches that provided results up to 7 attributes in much more time. Moreover, \textsc{KGen} demonstrates to preserve the quality of data correctly, a critical feature in order to keep the dataset qualitatively usable.


\textcolor{blue}{From a software and information systems engineering perspective the concrete usage} of our proposed method \textsc{KGen} is twofold: 
(a) privacy-aware data-intensive applications \cite{GuerrieroTRMBA17} \cite{GuerrieroTN18} could be designed using \textsc{KGen} as a middleware to anonymize datasets before processing automatically;
(b) compliance officers can use \textsc{KGen} to experiment with processed and non-processed data to quantify the extent of privacy ``damage'' carried out by data processors.
%

\textcolor{\editcolor}{The remaining part of the paper is organized as follows. Section \ref{sec:background} introduces the state of the art of the anonymization process and the main works related to anonymization. Sec. \ref{sec:kgen} introduces \textsc{KGen}, explaining all its components. Sec. \ref{sec:research_design} outlines the research design of the work. It describes the dataset used for the experimentation, the metrics used to evaluate the RQs illustrated above and the algorithms used for the comparison study. The results o this work are shown in Sec. \ref{sec:results}. Sec. \ref{sec:discussion} contains the discussion above the results obtained in the Sec. \ref{sec:results}. In Sec. \ref{sec:ttv} are discussed the threats to validity found in \textsc{KGen}. Lastly, section \ref{sec:conclusion} summarizes the main contributions of \textsc{KGen} and sketches future research directions.}

\section{Background and related work}\label{sec:background}

\begin{figure}[t]
    \centering
    \includegraphics[width = .8\textwidth]{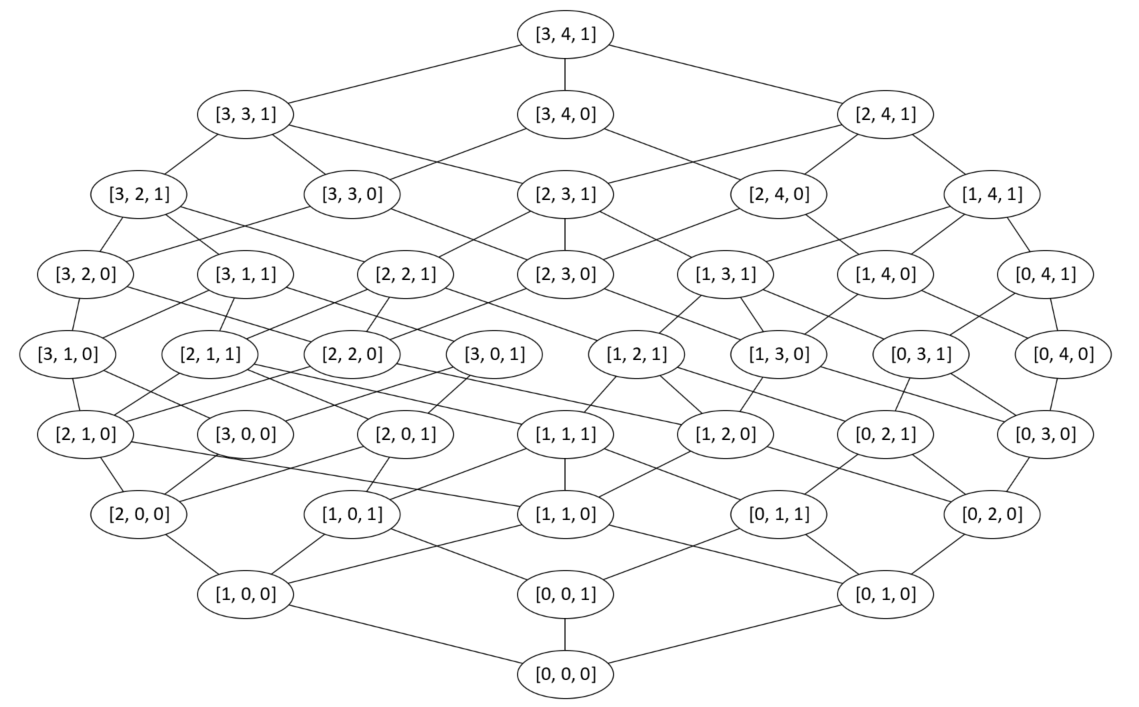}
    \caption{\footnotesize Example of lattice (Age-Postcode-Gender). Each node contain a possible level of generalization, for each attribute, and is connected to other nodes that can be reached increasing or decreasing by one a single level of generalization of a given node.}
    \label{fig:lattice}
\end{figure}

This section is organized in three main subsections: the first one describes the anonymization process to allow a better understanding of the purposes behind this work; the second subsection explains what a genetic algorithm is --- hence laying the technical foundations behind the metaheuristic underlying \textsc{KGen}. Third, finally, we showcase the known k-anonymity implementations in the state of the art to which \textsc{KGen} can be compared.


\subsection{Anonymization}
The anonymization process starts from a given dataset and generates an anonymous dataset. A dataset is composed of multiple observations with several different attributes. From a privacy perspective, there are two different kinds of attributes in any dataset \cite{samarati2001microdata}:

\begin{itemize}
    \item \textbf{Identifiers.} An Identifier attribute can uniquely identify a row in the dataset. In the anonymization process, these are suppressed (this process is explained more in-depth in the next section).
    \item \textbf{Quasi Identifiers.} Are the set of attributes that can be superimposed with external information to reveal an individual's identity \cite{dalenius1986finding}. Examples of common quasi-identifiers are \cite{el2009evaluating, el2006evaluating, el2007pan, canadian2005cihr}: dates (such as birth, death, admission, discharge, visit, and specimen collection), locations (such as postal codes, hospital names, and regions), race, ethnicity, languages spoken, aboriginal status, and gender.
\end{itemize}


During the anonymization process, the data is changed by either removing or suppressing all identifiers \cite{samarati2001microdata}. 
\textcolor{\editcolor}{This is essential to prevent reverting to the original dataset. Thus, nullifying the anonymization process}.
Stemming from this assumption, the only data that needs to be (partially)-anonymized while simultaneously ensuring the highest amount of information usability as possible are the quasi-identifiers.

Therefore, the central part of the anonymization process revolves around two main factors (1) the anonymization of those attributes, quasi-identifiers, and (2) finding the optimal trade-off between them. Hence, making it hard to uniquely identify rows in a data set by removing information and maximizing the usefulness of the data, keeping as much as possible intact. In turn, the usability of the dataset can be measured using the loss of information metrics \cite{el2009globally}. Metrics that are used to evaluate the goodness of a possible k-anonymous are explained below.

\subsection{K-Anonymity}

\begin{table}[!t]
\centering
\renewcommand{\arraystretch}{1.5}
\footnotesize
\caption{\footnotesize Original dataset. The attribute Name is an Identifier. Instead Age, Gender and Postcode are Quasi-Identifiers.}
\resizebox{.5\linewidth}{!}{
    \begin{tabular}{ccccc}
        \hline
        \textbf{Name} & \textbf{Age} & \textbf{Gender} & \textbf{Postcode} & \textbf{Crime} \\
        \hline
        Alice & 24 & F & 80015 & Assault \\
        Max & 28 & M & 80019 & Kidnapping \\
        Laurel & 42 & F & 85073 & Homicide  \\
        Frank & 49 & M & 85071 & Rape \\ 
        \hline
    \end{tabular}
}
\label{tab:dataset_table}
\end{table}

\begin{table}[!t]
    \centering
    \caption{\footnotesize Dataset k-anonymized. Considering the QI, the number of indistinguishable rows are two. So, the dataset is k-anonymized (k = 2).}
    \renewcommand{\arraystretch}{1.5}
    \resizebox{.5\linewidth}{!}{
    \begin{tabular}{ccccc}
        \hline
        Name & Age & Gender & Postcode & Crime \\
        \hline
        ***** & 20 - 30 & P & 8001* & Assault \\
        ***** & 20 - 30 & P & 8001* & Kidnapping \\
        ***** & 40 - 50 & P & 8507* & Homicide  \\
        ***** & 40 - 50 & P & 8507* & Rape \\
        \hline
    \end{tabular}
    }
    \label{tab:anonymized_dataset_table}
\end{table}

To guarantee anonymity \textsc{KGen} harnesses the concept of k-anonymity \cite{samarati2001microdata}. A dataset is called \textit{k-anonymous} if a single row is indistinguishable from, at least, other k-1 rows in the dataset.


\textit{\textbf{Definition:} Let $T(A_{1},...,A_{n})$ be a table and $QI_{T}(A_{1},...,A_{j})$ be all the quasi-identifiers of that table. T is said k-anonymous if, for each row of T, there are at least k-1 rows equals to that row (for a total of k indistinguishable rows).}



Table \ref{tab:anonymized_dataset_table} shows an example of anonymization of the dataset in Table \ref{tab:dataset_table}. The quasi-identifiers have been anonymized in order to guarantee the anonymization. Applying different levels of generalization for all quasi-identifier attributes, it is possible to guarantee the anonymization with a certain degree of remaining usability of the same dataset. Table \ref{tab:anonymized_dataset_table}, for example, shows a k-anonymous dataset with a level of k = 2. 


\subsection{\textcolor{\editcolor}{K-Anonymity operators}} \label{sec:generalization-suppression}




\begin{figure}[h]
   \begin{subfigure}[b]{\textwidth}
   \centering
   \includegraphics[width=0.7\textwidth, trim={0, 3cm, 0, 0}]{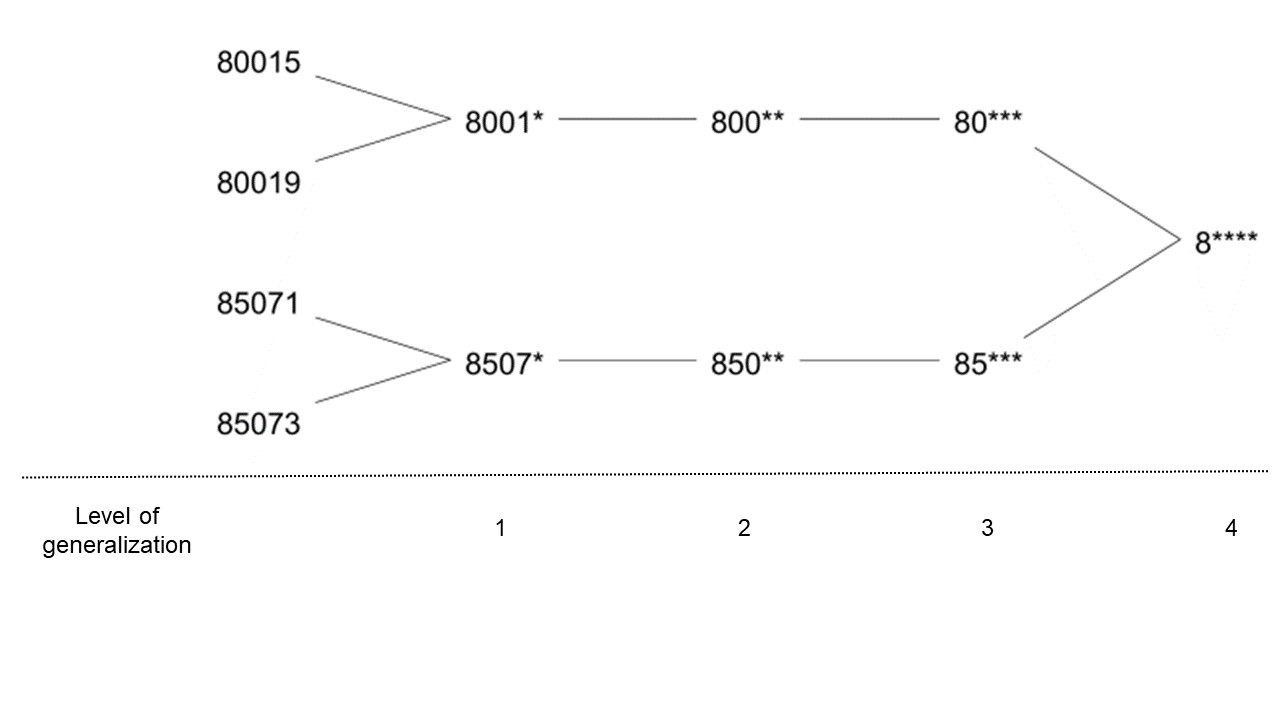}
   \caption{Generalization: POSTCODE}
   \label{fig:postcode} 
\end{subfigure}
\begin{subfigure}[b]{\textwidth}
   \centering
   \includegraphics[width=0.7\textwidth, trim={0, 3cm, 0, 0}]{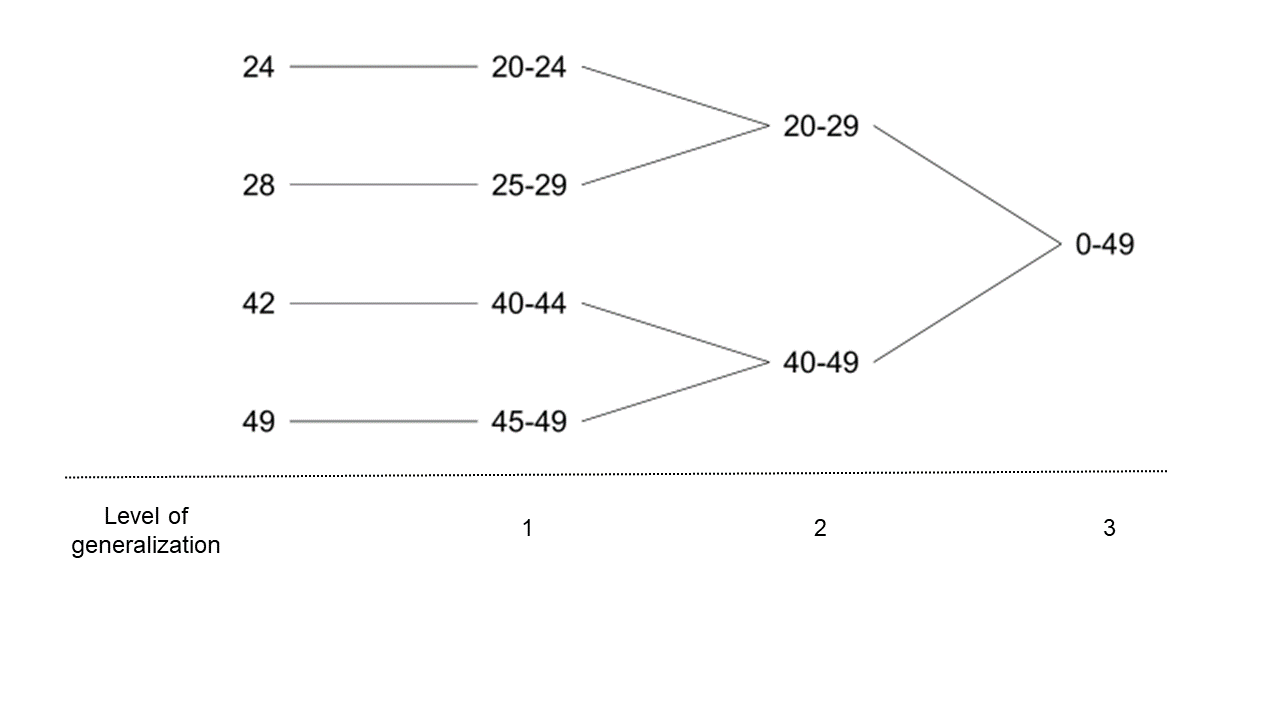}
   \caption{Generalization: AGE}
   \label{fig:age}
\end{subfigure}
\caption{\footnotesize Generalization hierarchy of two quasi-identifiers attributes.}
\label{fig:generalization}
\end{figure}

As mentioned before, the anonymization process revolves around the anonymization of attributes. State of the art offers several approaches, mainly around \textcolor{\editcolor}{four} different anonymization techniques, namely, \textcolor{\editcolor}{generalization, suppression, anatomization and perturbation} \cite{samarati2001microdata, fung2010privacy}.

\begin{itemize}
    \item \textbf{Generalization.} Given an attribute, its level of anonymity can be represented as a hierarchy (Fig. \ref{fig:generalization}).
    The higher the level of generalization of an attribute, the more the dataset is generalized, ensuring a high level of anonymization and a correspondingly low level of usability.

    \item \textbf{Suppression.} If a dataset is not k-anonymized because there is only a single row that does not allow to satisfy the k-anonymity conditions, it is possible to suppress that single row to have a k-anonymized dataset. 
    
    \item \textcolor{\editcolor}{\textbf{Anatomization.} Unlike generalization and suppression, the anatomization operator does not work on QI and sensitive data, but it works on the relationship between them. The operator splits the QI and the sensitive data into two different tables. To preserve the relationship between the two groups, each table have a common attribute, groupID, All rows in the same group have the same groupID \cite{fung2010privacy}.}
    
    \item \textcolor{\editcolor}{\textbf{Perturbation.} The perturbation replaces the original values with synthetic data. The new record generated does not correspond to a real-world record. In this way, for the attacker is not possible to recover sensitive data, starting from the data published.}
    
\end{itemize}

\textcolor{\editcolor}{\textsc{KGen} uses only generalization and suppression operators because, in the comparison study done in this work, the state-of-the-art approach chosen uses only the two operators mentioned above.}

Generalization works on the generalization of all values of a single attribute. Thus, no information is lost, but the entire dataset is modified. Conversely, suppression works at a local level, its approach revolving around the removal of entire rows, with the remaining data left unchanged \cite{samarati2001microdata}.

In both cases, however, it is always possible to compute the generalization hierarchy of all the attributes as represented by a lattice (i.e., repeating arrangement of points, see Fig. \ref{fig:lattice})\cite{el2009globally}. Thus, a node of the lattice represents a possible anonymized dataset containing the level of generalization of each \emph{quasi-identifier} attribute.  
The lattice shown in Fig. \ref{fig:lattice} is the representation of all possible configurations of the dataset in Tab.~\ref{tab:dataset_table}. 
The minimum node in a lattice is the representation of a dataset with all quasi-identifier attributes not anonymized (node $($000$)$ of Fig.~\ref{fig:lattice}); the maximum node, instead, is the representation of a dataset completely anonymized because contains the maximum level of generalization of each \emph{quasi-identifier} attribute (node $($341$)$ of Fig.~\ref{fig:lattice}).
Each arrow represents a possible generalization path taken through the lattice. Thus, the height of a lattice is equaled to the number of steps that, from the minimum node, are necessary to reach the maximum node, increasing one by one the level of generalization of a \emph{quasi-identifier} attribute.
Climb up the lattice allows to have a higher level of anonymization of a dataset but a lower utility (this concept is explained in Sec.~\ref{subsec:loss_of_information}).

Every path starting from the minimum node to the maximum node is called strategy path. For example, in the Fig. \ref{fig:lattice} the path [$($000$)$, $($001$)$, $($011$)$, $($021$)$, $($031$)$, $($041$)$, $($141$)$, $($241$)$, $($341$)$] is a strategy path.

All strategy paths share the same starting node (the minimum node of the lattice) and final node (the maximum node of the lattice). As explained before, since the maximum node represents a dataset completely anonymized, all strategy paths ensure the existence of at least one k-anonymized node.
3
In the lattice, every node could represent a k-anonymized dataset and, among these, only one represents the optimal global solution. So, the goal of k-anonymity is to find it in a reasonable time.


\subsection{Measuring Loss of information} \label{subsec:loss_of_information}

Using generalization and suppression, all possible datasets in the lattice can be possible solutions. The way of preferring a dataset to another for \textsc{KGen} is to select the dataset whose information is most useful in generalization. A dataset with more generalization or more suppression has less information and, hence, lower usability. \textsc{KGen} uses metrics to measure the usability of an input dataset using different metrics of information loss. The significant metrics for information loss are outlined below. Subsequently, a selection is made and illustrated for \textsc{KGen}.

One metric for the level of information loss was proposed by Samarati \cite{samarati2001microdata}. The idea of the proposed approach is to take the k-anonymity node with a minimum height level in the lattice. So, for example, if in the lattice showed in Fig. \ref{fig:lattice} nodes $($100$)$ and $($001$)$ are both k-anonymized, using this metric, they have the same level of loss of information because they have the same height level in the lattice.
However, the height lattice is not a helpful metric since it does not consider each attribute's maximum level of generalization. In the previous example, there are two nodes: the first one has only the first attribute generalized at level 1 of a maximum of 4 levels. Instead, the second one has the last attribute that, in this case, is completely anonymized. Moreover, with the first metric presented, they have the same level of loss of information.
Sweeney in \cite{sweeney2002achieving} and \cite{sweeney2001computational} takes into consideration as information metric also the level of generalization of each attribute. The aim is to evaluate, for each attribute, its level of generalization, called ``precision'', using this formula:

\textcolor{\editcolor}{
\begin{equation} \label{eqn:precision}
    Precision_{i} = \frac{log_{i}}{Hlog_{i}} \quad \forall i = 1,...,N
\end{equation}
}

\textcolor{\editcolor}{where log is the actual level of generalization of the i-th quasi-identifier, Hlog is the heigth of the generalization hierarchy of the i-th quasi-identifier and N is the total number of quasi-identifier attributes in the dataset.}
Hence, the level of generalization of a single node is given by the average of all precision values calculated.

\begin{equation}
    Precision = \frac{\sum_{i=1}^{N} Precision_{i}}{N}
\end{equation}
\label{eqn:prec}

For example, 
the node [1, 0, 0], 
\textcolor{\editcolor}{representation of the attributes Age/Postcode/Gender with a generalization hierarchy's height of, respectively, 3, 4 and 1,}
has a precision level of
\textcolor{\editcolor}{($\frac{1}{(3)}$ + $\frac{0}{(4)}$ + $\frac{0}{(1)}$) / 3 = 0.11.} 
Instead, the node [0, 0, 1] has a precision level of 
\textcolor{\editcolor}{($\frac{0}{(3)}$ + $\frac{0}{(4)}$ + $\frac{1}{(1)}$) / 3 = 0.33.}
With this metric, the node position in the lattice and the level of generalization of each attribute are taken into account. \textsc{KGen} uses this decaying information metric to find the dataset with the most information and the highest anonymization concurrently.

\subsection{K-Anonymity Complexity}

Different works prove that an optimal k-anonymization algorithm is an NP-Hard problem. Meyerson et al \cite{meyerson2004complexity} provide a demonstration on the complexity classification of the problem, finding that not only the k-anonymity algorithm is NP-Hard, but also the k-anonymization with suppression of different attributes is NP-Hard.

Aggarwal \cite{aggarwal2005k} shows that the k-anonymity complexity is highly dependent on the size of the problem and that it is impossible to apply the k-anonymization property on a dataset with lots of quasi-identifier attributes with an acceptable level of information loss.

Sun et al. \cite{sun2008complexity} introduce two variants of the k-anonymization problem, the Restricted K-anonymity problem and the Restricted K-anonymity problem on attributes. They proved that both of them are \textit{NP}-Hard for $k \geq 3$, but, on the positive side, they developed a polynomial solution for the k-anonymization problem with $k=2$.

\subsection{Genetic Algorithms: An Overview}

Genetic algorithms are simulations of natural selection, used to solve optimization problems \cite{dansimon2013ga} such as the one reflected by \textsc{KGen}. The natural selection process inspires genetic algorithms, and their workings and architecture reflect the natural process of reproduction, proliferation, and selection. More specifically, starting from an initial population, the algorithm selects, with a function used to measure the goodness of an individual, the best individuals and, from them, produces new individuals. Then, the old and the new population are re-evaluated to see which of them survives to the next generation. This process goes on until a stop condition is satisfied.
In order to better explain this process, it is essential to describe \textcolor{\editcolor}{the main components of a genetic algorithm}:
\\\\
\textcolor{\editcolor}{\textbf{Solution encoding:} a good solution representation plays a key role in a genetic algorithm because all future evaluations are applied to all solutions. So, if a solution is easy to evaluate, then the entire algorithm's complexity is low. A solution typically consists in an array of values. As a first step, a random population is generated. Then the algorithm tries to improve its solutions in order to find the best solution.} 
\\\\
\textcolor{\editcolor}{\textbf{Fitness function:} in implementing a genetic algorithm, a key role is played by the complexity of the fitness function. A fitness function is a good representation of the objective to achieve. If it has low complexity, then the entire algorithm has a lower complexity. The choice of the proper fitness function should be made together with the choice on the solution encoding because they are highly correlated. The fitness function is directly applied to the solution, so if they are incompatible, then the evaluation process is more complicated.}
\\\\
\textcolor{\editcolor}{\textbf{Genetic operator:} Genetic operators are functions that automatically allow the generation of new chromosomes, starting from the previous population. There are three different types of operators: selection, an operator used to find the best chromosome in the population; crossover, a "mating process" applied to two chromosomes to generate two new chromosomes; mutation, operator used to mutate a single chromosome to avoid the genetic algorithm convergence into a local optimal solution \cite{dansimon2013ga}.}
\\\\

\subsection{Related work} \label{subsec:related}

There are many works on k-anonymization and its practical implementation.
Samarati et al. \cite{samarati2001microdata} provide a k-minimal generalization algorithm to apply a binary search to find all k-anonymous node, selecting all nodes with the least steps as solutions. If there is more than one node as a solution, the algorithm selects one randomly or using other criteria, as the information loss. However, the node with the lowest distance vector is not guaranteed the optimal solution because they could be other nodes with a higher distance value but with a lower level of information loss. For this reason, the algorithm does not provide the optimal global solution.

Similarly, the \emph{Datafly} algorithm adopts a heuristic based on the attribute \cite{sweeney1997guaranteeing}\cite{sweeney2002achieving}. The most distinct attribute is taken into account as how next generalized attribute. The process continues with new distinct attributes that do not satisfy k-anonymous until the k-anonymous criteria are satisfied. This approach does not guarantee the minimum k-anonymous solution, however, the found solution is always k-anonymous.

Kirsten et al. \emph{Incognito} exploits a bottom-up approach with a breadth-first strategy to navigate the lattice to find all k-minimal distance vectors \cite{lefevre2005incognito}. After detecting all vectors, the algorithm calculates their information loss to select the solution with the least information loss as the optimal solution. This algorithm can find, in this way, a global optimum.

Besides, the \emph{Optimal Lattice Anonymization (OLA)} The OLA algorithm is an improvement of Incognito and Datafly algorithms \cite{el2009globally}. All the anonymization processes, as shown in Fig. \ref{fig:lattice}, may be represented as a lattice. The goal of the OLA algorithm is to find the optimal node in the lattice that must be k-anonymous and with minimum loss of information. The approach embraces a binary search algorithm for each strategy path. When the optimal node in a strategy path is reached, the algorithm commences to analyze the next strategy hub, and so on. In the end, the algorithm holds a list with all k-minimal nodes for each strategy path. At this point, it is chosen only the node with the minimum information loss. Thus, OLA, as Incognito, can provide a globally optimal solution.

Bayardo et al. \cite{bayardo2005data} present a new approach to explore the space of possible combinations developing data-management strategies to reduce reliance on expensive operations. They can find an optimal solution under two representative cost measures and a wide range of k. Moreover, they can provide good anonymizations where the input data or input parameters preclude finding an optimal solution in a reasonable time.

Lyengar shows an example of a Genetic algorithm applied on the k-anonymity problem \cite{iyengar2002transforming}. It seems to generate good results, as we can see from the experimentation done in their work. Nevertheless, they considered only a dataset with eight quasi-identifier attributes, lacking more considerable experimentation.

Among all of these k-anonymization algorithms, only OLA and Bayardo's algorithm proved that their results are better than the others (Datafly, Samarati's algorithm) \cite{el2009globally,bayardo2005data}.
For this work, we realized a comparison only with OLA because we found different implementations of it, differently from Bayardo's approach.
Furthermore, we did not realize a comparison with Lyengar's GA because of lacking a pseudo-code of the algorithm or a repository with their work.

\section{Scalable K-Anonymization: \textsc{KGen} Explained} \label{sec:kgen}

\begin{figure}[t]
    \centering
    \includegraphics[width = \textwidth]{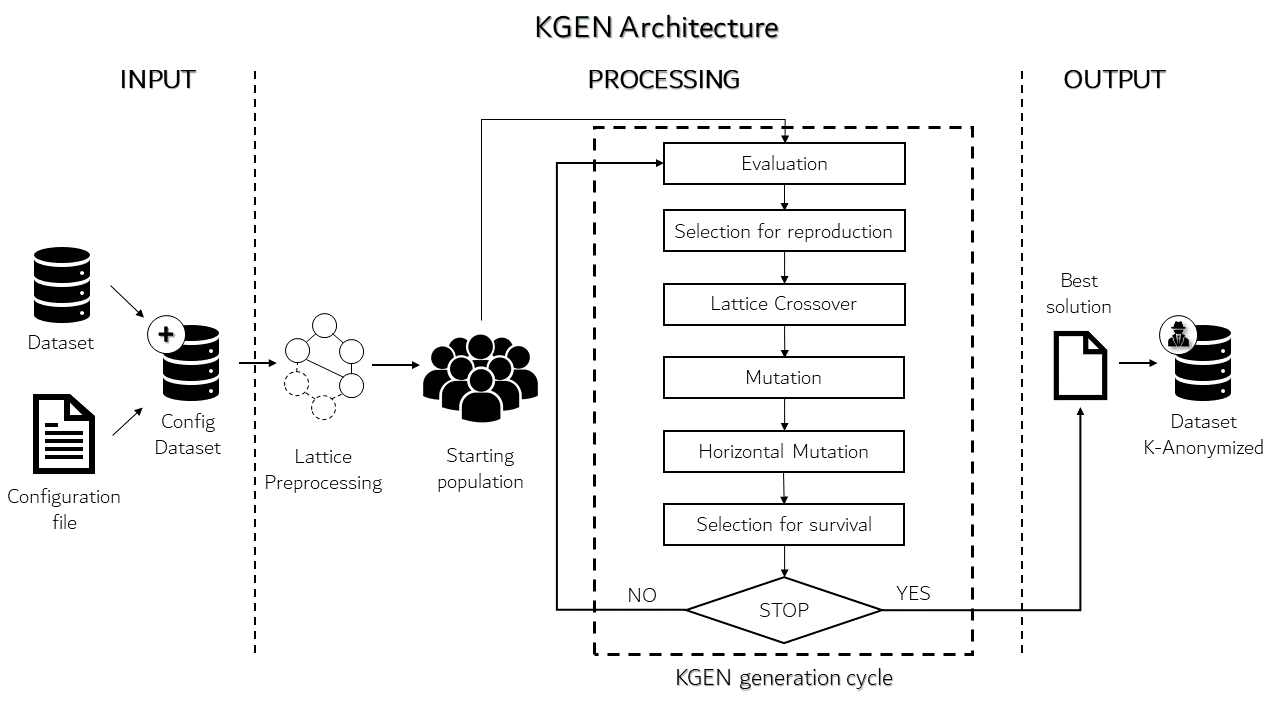}
    \caption{\footnotesize KGEN Pipeline. It is divided into three steps (separated bu dotted vertical lines), input, processing, and output; the KGEN-GA architecture is described in the processing step.}
    \label{fig:kgen_pipeline}
\end{figure}

This section describes \textsc{KGen} from a technical perspective, elaborating (1) the general \textsc{KGen} architecture; (2) the \textsc{KGen} lattice preprocessing; (3) solution encoding; (4) solution fitness; (5) genetic operators. 

\subsection{\textsc{KGen} Architecture}
An overview of \textsc{KGen} architecture is shown in Fig. \ref{fig:kgen_pipeline}. Processing of data starts with an input phase in which \textsc{KGen} receives a dataset to anonymize along with configuration parameters such: (a) the generalization strategy to be adopted; (b) attributes' information type, that is, whether they are Identifiers or Quasi-Identifiers.
As explained by Samarati et al.~\cite{samarati2001microdata}, there are different generalization strategies, assuming the existence of different domains, including generalized values and mapping between each domain and domains generalization of it. Thus, for example, the postcode can be generalized, dropping, from the right, the least significant value (as shown in Fig.~\ref{fig:postcode}).

The subsequent processing phase is the core of the \textsc{KGen} approach. An overview of this phase is provided in Algorithm~\ref{alg: kgen_pseudocode}. The first step of \textsc{KGen} processing phase is the preprocessing of the lattice for size reduction. The next step is an iteration of the \textsc{KGen} Genetic Algorithm (GA) implementation. In the \textsc{KGen}-GA step, \textsc{KGen} tries to converge to the optimal solution following the GA meta-heuristic approach recapped in Section \ref{sec:background}. 
The output of the processing phase is the k-anonymized dataset using the best solution provided by \textsc{KGen}. 

{\footnotesize
\begin{algorithm}[h]
    \caption{KGEN Algorithm}
    \label{alg: kgen_pseudocode}
    \hspace*{\algorithmicindent} \textbf{Input:} \textit{Dataset} \\
    \hspace*{\algorithmicindent} \textbf{Output:} \textit{Dataset anonymized}
    \begin{algorithmic}[1]
        \Procedure{KGEN Algorithm}{}
        \State $bounds \gets LatticePreprocessing(dataset)$
        \State $t \gets 0$
        \State $P_{t} \gets initRandomPopulation(bounds)$
        \State $evaluate(P_{t})$
        
        \While {$evaluation < maxEvaluations$}
            \State $\textit{Offsprings} \gets \textit{empty offspring list}$
            \For {$(i = 0; i < populationSize; i+=2)$}
                \State $parents \gets selection(Population)$
                \State $\textit{tmpOffsprings} \gets crossover(parents)$
                \State $mutation(\textit{tmpOffsprings})$
                \State $horizontalMutation(\textit{tmpOffsprings})$
                \State $\textit{Offsprings}.add(\textit{tmpOffsprings})$
                \State $evaluation = evaluation + 2$
            \EndFor
            \State $evaluate(\textit{Offsprings})$
            \State $P_{t} \gets P_{t} \cup \textit{Offsprings}$
            \State $P_{t+1} \gets selection(P_{t})$
            \State $t \gets t+1$
        \EndWhile
        
        \State $S \gets minLOGSolution(P_{t})$
        \State $newDataset \gets anonymize(dataset, S)$ \\
        \Return $newDataset$
        
        \EndProcedure
    \end{algorithmic}
\end{algorithm}}

\subsection{Lattice Preprocessing} \label{subsec:lattice_preprocessing}

\textcolor{\editcolor}{The lattice reduction is the first step of \textsc{KGen} execution. It is based on the lattice pruning technique used in \cite{lefevre2005incognito}.}
This step aims at removing the complexity given by the generation of a lattice at the expense of introducing an acceptable permutation computational cost.
\textcolor{\editcolor}{It} reduces the lattice size, thus the complexity of the k-anonymization algorithm.
The size-reduction process exemplified in Fig. \ref{fig:lattice} shows an example of a non-reduced lattice. In this example, the minimum node is $<$0, 0, 0$>$ and the maximum node is \textcolor{\editcolor}{$<$3, 4, 1$>$}. The reduction technique is recapped in Table \ref{tab:lattice_reduction_table}, parts from (a) to (f); \textsc{KGen} slices the dataset into N vectors, one per quasi-identifier (Tab. \ref{tab:lattice_preprocess_first_step}), and validates the k-anonymity property iteratively on each vector thus obtained, until a new minimum level of generalization is found (Tab. \ref{tab:lattice_preprocess_second_step}).
\textcolor{\editcolor}{The idea is that if at least one quasi-identifier attribute is not k-anonymized, then the entire dataset cannot be anonymized too. Hence, the computational cost for the execution of \textsc{KGen} on nodes containing quasi-identifiers not anonymized is meaningless.}
Although this approach poses limitations when anonymizing by suppression, such limitations are addressed in the Threats to Validity section, see Sec. \ref{sec:ttv}.

\begin{table}[!ht]
    \footnotesize
    \renewcommand{\arraystretch}{1.5}
    
    \caption{\footnotesize Example of the entire lattice reduction process.}
    \begin{subtable}{\linewidth}
        \centering
        \begin{tabular}{ccc}
            \hline
            \textbf{Age} & \textbf{Postcode} & \textbf{Gender}\\
            \hline
            24 & 80015 & F \\
            28 & 80019 & M \\
            42 & 85073 & F \\
            49 & 85071 & M \\ 
            \hline
        \end{tabular}
        \caption{\footnotesize Original dataset not anonymized. The attributes Age, Gender and Postcode are Quasi-Identifiers.}
        \label{tab:lattice_preprocess_original_dataset}
    \end{subtable}
    
    \begin{subfigure}{\linewidth}
        \begin{subtable}{.33\linewidth}
          \centering
            \begin{tabular}{c}
                \hline
                \textbf{Age} \\
                \hline
                24 \\
                28 \\
                42 \\
                49 \\ 
                \hline
            \end{tabular}
        \end{subtable}%
        \begin{subtable}{.33\linewidth}
          \centering
            \begin{tabular}{c}
                \hline
                \textbf{Postcode} \\
                \hline
                80015 \\
                80019 \\
                85073 \\
                85071 \\ 
                \hline
            \end{tabular}
        \end{subtable} 
        \begin{subtable}{.33\linewidth}
          \centering
            \begin{tabular}{c}
                \hline
                \textbf{Gender} \\
                \hline
                F \\
                M \\
                F \\
                M \\
                \hline
            \end{tabular}
        \end{subtable}
        \caption{\footnotesize First step of the reduction process. The original dataset is split into n dataset, where n is the number of quasi-identifiers in the original dataset. Each of these new datasets contain only one of these quasi-identifiers.}
        \label{tab:lattice_preprocess_first_step}
    \end{subfigure}
    \begin{subfigure}{\linewidth}
        \begin{subtable}{.33\linewidth}
          \centering
            \begin{tabular}{c}
                \hline
                \textbf{Age} \\
                \hline
                20 - 29 \\
                20 - 29 \\
                40 - 49 \\
                40 - 49 \\ 
                \hline
            \end{tabular}
            \caption{Age: LOG 2.}
        \end{subtable}%
        \begin{subtable}{.33\linewidth}
          \centering
            \begin{tabular}{c}
                \hline
                \textbf{Postcode} \\
                \hline
                8001* \\
                8001* \\
                8507* \\
                8507* \\ 
                \hline
            \end{tabular}
            \caption{PC: LOG 1.}
        \end{subtable} 
        \begin{subtable}{.33\linewidth}
          \centering
            \begin{tabular}{c}
                \hline
                \textbf{Gender} \\
                \hline
                F \\
                M \\
                F \\
                M \\
                \hline
            \end{tabular}
            \caption{Gender: LOG 0}
        \end{subtable}
        \caption{\footnotesize Second step of the reduction process. Each of the datasets generates previously has been anonymized up to reach the minimum level of anonymization. The level of generalization of each of these datasets represent the new minimum level of generalization of the lattice.}
        \label{tab:lattice_preprocess_second_step}
    \end{subfigure}
    
    \label{tab:lattice_reduction_table}
\end{table}

\subsection{Solution Encoding}

\begin{figure}[!h]
    \centering
    \includegraphics[width = .5
    \textwidth]{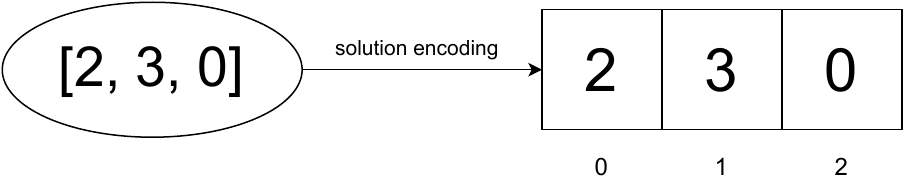}
    \caption{\footnotesize Solution encoding of the lattice node $<2, 3, 0>$.}
    \label{fig:solution-encoding}
\end{figure}

A genetic algorithm aims to find the best pseudo-optimal solution in a reasonable time. In this case, a solution is the representation of a node in the lattice  (see Fig. \ref{fig:lattice}) that represents its level of generalization.
In \textsc{KGen}, a solution is represented as an array of numbers, \textcolor{\editcolor}{where in the i-th position of the array contains the value of the i-th attribute in the lattice node. Fig. \ref{fig:solution-encoding} shows the solution encoding of the lattice node Age/Postcode/Gender $<$2, 3, 0$>$. In the solution encoding process, the level of generalization values of Age, Postcode and Gender are respectively put in positions 0, 1 and 2.}
In a Genetic algorithm, the initial population is initialized randomly.

\subsection{Fitness Functions}
Every Genetic Algorithm needs to define its fitness function. This function allows evaluating, for each iteration, all generated solutions. As discussed in section \ref{sec:background}, there are two metrics for the evaluation of a single node, namely, (a) k-anonymity and (b) loss of information. In \textsc{KGen}, the loss of information is the only metric used to evaluate the fitness of a solution. For every fitting solution, k-anonymity is evaluated to see if a solution is feasible or not. Thus, the goal of the \textsc{KGen} fitness function is to find the lowest value of loss of information of a node while ensuring, at the same time, the k-anonymity property.

\begin{table}
\centering
\footnotesize
    \renewcommand{\arraystretch}{1.5}
    \caption{\footnotesize Example of support map for the quasi-identifiers AGE and Postcode.}
    \begin{subtable}{.5\linewidth}
      \centering
        \begin{tabular}{ccc}
            \hline
            \textbf{Value} & \textbf{LOG} & \textbf{Rows} \\
            \hline
            24 & 0 & [1] \\
            28 & 0 & [2] \\
            42 & 0 & [3] \\
            49 & 0 & [4] \\
            20-24 & 1 & [1] \\
            25-29 & 1 & [2] \\
            40-44 & 1 & [3] \\
            45-49 & 1 & [4] \\
            20-29 & 2 & [1, 2] \\
            40-49 & 2 & [3, 4] \\
            0-49 & 3 & [1, 2, 3, 4] \\
            0-99 & 4 & [1, 2, 3, 4] \\
            \hline
        \end{tabular}
        \caption{Age support map.}
        \label{tab:age_support_map}
    \end{subtable}%
    \begin{subtable}{.5\linewidth}
      \centering
        \begin{tabular}{ccc}
            \hline
            \textbf{Value} & \textbf{LOG} & \textbf{Rows} \\
            \hline
            80015 & 0 & [1] \\
            80019 & 0 & [2] \\
            85073 & 0 & [3] \\
            85071 & 0 & [4] \\
            8001* & 1 & [1, 2] \\
            8507* & 1 & [3, 4] \\
            800** & 2 & [1, 2] \\
            850** & 2 & [3, 4] \\
            80*** & 3 & [1, 2] \\
            85*** & 3 & [3, 4] \\
            8**** & 4 & [1, 2, 3, 4] \\
            ***** & 5 & [1, 2, 3, 4] \\
            \hline
        \end{tabular}
        \caption{Postcode support map}
        \label{tab:postcode_support_map}
    \end{subtable}
    \label{tab:support_map_table}
\end{table}

\subsubsection{Implementing K-Anonymity in \textsc{KGen}} 
We implemented \textsc{KGen} using the \textcolor{\editcolor}{improved algorithm for k-anonymity} presented by Zhang et al. \cite{zhang2012improved}. 
\textcolor{\editcolor}{They propose a technique for improving the k-anonymity implementation} 
by providing a new structure for the generalization hierarchy, namely, a \emph{support map}. A support map provides a structure in which each indistinguishable value is associated with its level of generalization, and all the rows contain an equal value.
Tab. \ref{tab:support_map_table} shows an example of a support map, applied on two quasi-identifier attributes, Age and Postcode. With the support \textcolor{\editcolor}{map} technique, for each attribute, there is a related support map. This support map contains all values referred to that attribute, including all their generalization versions, and, for each value, they memorize its level of generalization and all rows that contain that value. In Tab \ref{tab:age_support_map}, the value 24 has a level of generalization 0 and is included only in the first row. Differently, its generalization 20-29 has a level of generalization 2 and can be found in rows 1 and 2. In this way, to see if a dataset is k-anonymized, the algorithm intersects all value rows of a given level of generalization to see if there are no rows less than k. In Tab. \ref{tab:support_map_table}, for example, with the intersection of LOG 2 of age and LOG 1 of Postcode, we have two groups of rows: the first one containing rows 1 and 2, that contain values 20-29 and 8001*; the last one, that contains rows 3 and 4 with values 40-49 and 8507*.

\subsubsection{Implementing Loss of Information in \textsc{KGen}}

As discussed in Section \ref{subsec:loss_of_information}, \textsc{KGen} implements the precision criterion, as information loss metric. Each possible solution is evaluated with the precision Formula \ref{eqn:precision}.  The goal of \textsc{KGen}'s genetic algorithm is to minimize the precision of a solution to find the best k-anonymized solution with the least precision.

\subsection{Genetic Operators}

    

\begin{figure}[!h]
    \centering
    \renewcommand{\arraystretch}{1.5}
    \resizebox{.45\linewidth}{!}{%
        \begin{tabular}{cccc}
            \hline
            LOG & Penalty & LOG weighted & Percentage \\
            \hline
            0.85 & 0 & (1-0) * (0.85) = 0.85 & 0.51 \\
            0.44 & 3 & (1-0.3) * (0.44) = 0.308 & 0.19 \\
            0.55 & 1 & (1-0.1) * (0.55) = 0.495 & 0.3 \\
            \hline
        \end{tabular}
    }
    \qquad
    \resizebox{.45\linewidth}{!}{%
        \begin{tabular}{c}
             \\
            \begin{tikzpicture}
                \pie [color={black!10, black!30, black!50}]{51/LOG=0.85, 19/LOG=0.44, 30/LOG=0.55}
            \end{tikzpicture} \\
             \\
        \end{tabular}
    }
    \captionlistentry[table]{A table beside a figure}
    \caption{\footnotesize Selection process, based on LOG metric as fitness function. Based on their LOG value and their penalty, the selection generates all probabilities. The pie chart shows the probability to choose a single solution.}
    \label{fig:selection_log}
  \end{figure}
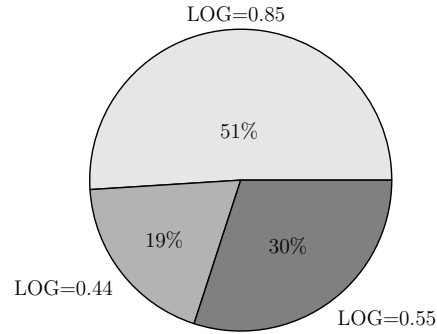

For the implementation of the \textsc{KGen}-GA approach, the following operators are provided.

\textbf{Selection.} For the selection operator, \textsc{KGen} uses the \textit{Tournament Selection} operator \cite{blickle1995mathematical} with penalty. 
The Tournament Selection is used to select the fittest candidate for the current generation. This operator assigns a probability to each solution based on two criteria: the fitness value and the penalty of a solution. The fitness value, in our case, is the loss of information metric. Instead, the penalty is calculated as follows: when a new solution is generated, its penalty value is $0$. Suppose this solution survives going to the next generation, its penalty increases by $1$. The maximum value reachable is $9$. Otherwise, with a value of $10$, the penalty decreases the probability to $0$. The concept is that the more a solution survives, the more the probability to be chosen decreases. Therefore, the penalty is used as a weight for solution optimality. An example of this process is shown in Fig. \ref{fig:selection_log}
\textcolor{\editcolor}{(in the figure, the data regarding the level of generalization (LOG) and the penalty are chosen randomly, just to explain the process behind the \textsc{KGen} selection operator).}
The probability of selection is calculated using this formula:

\begin{equation} \label{eqn:selection_prob}
    P(S_{j}) = \frac{log(S_{j}) * w_{j}}{\sum_{i=1}^{n}(log(S_{i}) * w_{i})}
\end{equation}

\begin{figure}[h!]
\centering
   \begin{subfigure}[h]{.35\textwidth}
   \includegraphics[width=\textwidth]{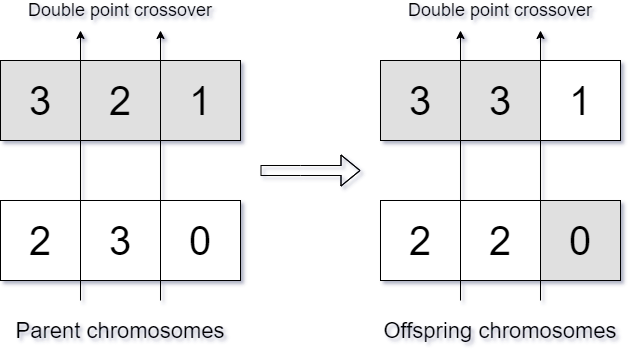}
   \caption{Offspring generation.}
   \label{fig:crossover_offspring} 
\end{subfigure}
\begin{subfigure}[h]{.6\textwidth}
   \includegraphics[width=\textwidth]{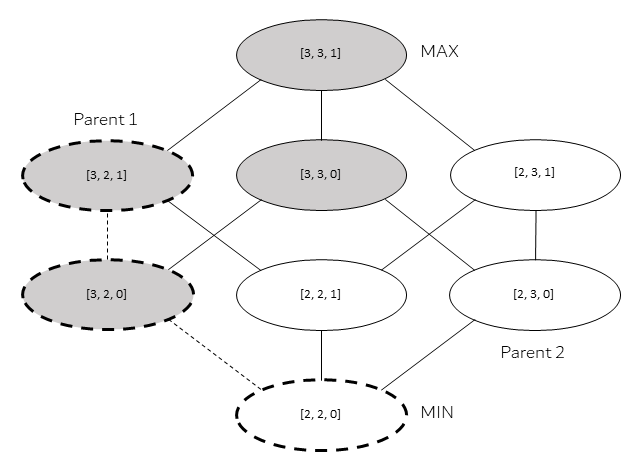}
   \caption{Crossover lattice, an example with only one k-anonymized parent.}
   \label{fig:crossover_lattice}
\end{subfigure}
\caption{\footnotesize Example of crossover operator with only one of the parent k-anonymized. In this case, all nodes with dashed lines represent a possible final offspring of the crossover.}
\label{fig:crossover}
\end{figure}

\textbf{Crossover.} \textsc{KGen} provides its own Crossover implementation, based on the double point crossover defined in \cite{mirjalili2019genetic}. 
Fig. \ref{fig:crossover_offspring} show the first step:
(i) the PARENTS selected with the selection operation, (ii) on top of them the crossover generates two new chromosomes, one with the highest value extracted from PARENTS and the second one with the lowest values extracted from PARENTS.\\
Subsequently, three possible scenarios manifest:

\begin{itemize}
    \item[-] \textbf{Case 1.} Both parents are k-anonymized. In this case, the maximum node is anonymized because, by definition of strategy path, all nodes after a k-anonymized node are also k-anonymized. If also the minimum node is anonymized, add it to the final offspring. Otherwise, the algorithm adds a random node between the minimum node and the first parent node and another random node between the minimum node and the second parent node;
    \item[-] \textbf{Case 2.} Both parents are not k-anonymized. In this case, the minimum node is not k-anonymized, and the final offspring is the maximum node;
    \item[-] \textbf{Case 3.} Only one of parents is k-anonymized. The minimum node is not k-anonymized, and the maximum node is k-anonymized. In this case, the last offspring is a random node between the minimum node and the k-anonymized parent. 
\end{itemize}

An example of case 3 is shown in Fig. \ref{fig:crossover}, while Fig. \ref{fig:crossover_offspring} shows the generation of the minimum and maximum nodes. Finally, Fig. \ref{fig:crossover_lattice} shows the crossover lattice that contains all the possible crossover's offspring. In this case, only nodes with dashed lines are considered since they represent the random solution discussed previously.

\textbf{Mutation.} In this case, \textsc{KGen} uses two different Mutation techniques:
\begin{itemize}
    \item[-] \textbf{Standard mutation.} a classic mutation operator, inherited from the approach in \cite{goldberg1988genetic}. This approach changes a single value of the chromosome and allows to change a possible solution with another one from the same strategy path. This operator needs to guarantee the principle of exploitation \cite{mkaouer2014model} since this principle allows a solution to move up or down its strategy path;
    \item[-] \textbf{Horizontal mutation.} this operator allows the genetic algorithm to change a solution with another solution of a different strategy path. In this way, it is possible to guarantee the exploration criteria. In order to change the strategy path, it is necessary to change more than one value of the solution and, to avoid having a solution in the same strategy path, it is necessary, alternatively, increase and decrease the chosen value, with a value between the minimum value (or maximum value in case we need to increase the value) and the actual node. An example of Horizontal mutation is shown below:\\ 

    \textbf{Example}\\
    \textit{Minimum solution:} 0 0 0 0 0 \\
    \textit{Actual solution:} 2 2 2 2 2 \\
    \textit{Maximum solution:} 4 4 4 4 4 \\
    \textit{Percentage of values to mutate:} 50\%. In this case it means that we need to mutate 2 values \\
    \textit{Random indexes chosen:} 2, 3 \\
    \textit{Algorithm:} The value in the index 2 can choose a random value between its value and its maximum (so, from 2 to 4). The value in the index 3 can choose, instead, a value between 2 and 0, its minimum. \\
    \textit{Possible mutate solution:} 2 2 3 0 2 \\
    This procedure of increasing and decreasing iteratively must keeps going on until all indexes chosen have been mutated.
\end{itemize}


\section{Research Design}\label{sec:research_design}
The main goal of this work is to provide an approach to the stakeholders that can be used in a real case scenario. To that end, we proposed \textsc{KGen}, a meta-heuristic approach based on a Genetic Algorithm, to build an infrastructure capable of anonymizing a dataset in a real case scenario. First, it means that the dataset specification can not know a priori, so the approach should scale with the dataset provided. Secondly, we evaluated the algorithm proposed with experimentation, using a large dataset to validate the approach in a significant case context.

\subsection{Dataset}
To answer the first main research question, we build an experimentation on top of the dataset provided by the Financial Forensics (F$^{2}$) Taskforce West-Brabant-Zeeland. The task force needed a middleware capable of enabling forensic analysis without putting at risk the privacy of data owners and without any human intervention over the data; furthermore, this needed to be done in computational times which were consistent with the \emph{quantity} of data available as opposed to the \emph{qualities} of that data. The task force has many instances of data constrained around a reasonable set of 50+ features. Therefore, the key requirement was striking a balance between the computational complexity of the algorithms involved and the anonymization reliability of such algorithms. We were provided with an experimental dataset in the scope of our experimentation, which was completely spoofed at the source. Namely, the data was disguised as a communication from an unknown source but still reflecting the original structure and properties. The dataset in question contained 47 attributes and 1599 observations involving four different attribute types: Dates, Numbers, Strings, Places. The generalization techniques used to generalize them are showed in Tab. \ref{tab:generalization_attributes}.

To validate \textsc{KGen} with a large dataset, we led a second experimentation using the ``c2k\_data\_comma.csv'' dataset \cite{cargo2000dataset}, which is commonly considered \emph{big} data (in terms of attributes, or columns of the dataset) for anonymization research, with its 97 attributes and 3942 observations. The attributes analyzed are all numeric, so the only generalization strategy applicable is the range generalization \cite{samarati2001microdata}. The more the range of possible values increases, the more a number is generalized (e.g., 23, at the level of generalization $1$ can be generalized in 20-25).

\begin{table}[h]
\footnotesize
    \centering
    \renewcommand{\arraystretch}{1.5}
    \caption{\textit{Generalization strategies applied on the K-Anonymity problem.}}
    \resizebox{.6\linewidth}{!}{
        \begin{tabular}{ll}
            \hline
            \multicolumn{2}{l}{Generalization tecniques} \\
            \hline
            NUMBER & Range generalization (3 -\textgreater 0-5) \\
            STRING & Star generalization (NL805 -\textgreater NL80*) \\
            DATE & Date generalization \\
             & (01/01/1970 -\textgreater 01/1970 -\textgreater 1970) \\
            PLACE & Place generalization \\
             & (Den Bosch -\textgreater Noord Brabant) \\
            \hline
        \end{tabular}
    }
    \label{tab:generalization_attributes}
\end{table}

\subsection{Metrics}

To find an answer to our minor RQ$s$ outlined in Sec. \ref{sec:introduction}, we defined the evaluation metrics below.

The RQ$_1$ compares the performance of the approach using execution time of the anonymization algorithm concerning the complexity of the dataset in input, as defined in related work \cite{el2009globally}. 
K-Anonymity property is an NP-Hard problem \cite{meyerson2004complexity}. For this reason, when the number of quasi-identifier attributes increases, the number of nodes in the lattice increases and, consequently, the execution time to analyze them. Hence, the execution time is a reliable indicator to compare approaches.

To answer to the RQ$_2$, we proposed a measure of accuracy, expressed as the distance between the optimal solution and pseudo-optimal solution. Each solution is part of a strategy path, and there is an optimal solution for each strategy path. Following this principle, the worst solution is the last node of this strategy path, with an accuracy value equal to 0. Instead, the optimal node has an accuracy value equal to 1. More in general, the accuracy of a solution is computed as follows:

\begin{equation} \label{eqn:accuracy}
    acc_{i} = 1 - \frac{\textcolor{\editcolor}{\mid H(S_{i}) - H(optS) \mid}}{H(worstS) - H(optS)}
\end{equation}

where H(x) is the height function of an x solution.
The general accuracy, instead, is the weighted arithmetic mean of all accuracy values of our solutions, formally:

\begin{equation} \label{eqn:accuracy}
    accuracy = \frac{\sum_{i=0}^{n}(\omega_{i}*acc_{i})}{\sum_{i=0}^{n}\omega_{i}}
\end{equation}

We choose the weighted arithmetic mean because of the $0$ value problem \cite{WOOD20061326}; in our case, accuracy could be $0$, and it is not possible to use harmonic or geometric means with values less or equal to $0$.
The problem with these metrics is that we should always know the optimal solution to measure the accuracy level. So, the only way to determine the accuracy level is to compare an approach with another one that provides optimal solutions.

In the RQ$_3$, we measure the quality of a proposed solution. 
The quality is strongly related to the anonymization and usability of a dataset. As previously stated, the metrics used to evaluate these two aspects are the level of generalization and the percentage of a solution's suppression. With the former, we measure the level of generalization of a solution, and the latter is used as an indicator of the level of suppression of a dataset. All solutions provided by an approach are k-anonymized. Therefore, the lower is the level of generalization and the level of suppression of a solution, the better its quality. Since solutions could be more than one, the final level of generalization is the minimum of all levels of generalizations of solutions and the level of suppression is taken from the solution found.

\subsection{Evaluated Algorithms}
In the scope of our evaluation, we select four k-anonymization algorithms from state of the art, which use generalization and suppression techniques as well as an exhaustive algorithm featuring a brute-force approach by enumeration \cite{Ullmann1976}. Below are listed the selected algorithms:

\begin{itemize}
    \item[--] \textbf{Exhaustive Approach.} This algorithm is an implementation of the k-anonymization property assessment algorithm as well as the generalization and suppression metrics on all nodes in the input lattice. After the analysis of the entire lattice, it is possible to find the minimum k-anonymization node. This approach provides the optimal solution;
    \item[--] \textbf{OLA Approach.} As explained in the Related Work section (see Sec. \ref{subsec:related}), the OLA algorithm is an optimization of the k-anonymization algorithm. Also, this algorithm converges towards the optimal solution;
    \item[--] \textbf{\textsc{KGen} Approach.} \textsc{KGen} is the approach that we want to test within this work, designed to cope with big datasets;
    \item[--] \textbf{Random-Search Approach.} This algorithm is included as a validation baseline for \textsc{KGen}. The comparison with this algorithm is due to genetic algorithms' feature of introducing a certain degree of randomness in solution generation. Hence, by comparing \textsc{KGen} to a Random algorithm, we aim at establishing whether the \textsc{KGen} behavior is close or not to a Random approach.
\end{itemize}

The remaining approaches from state of the art discussed in Sec. \ref{subsec:related} were already compared in other previous works with the OLA approach \cite{el2009globally}. For this reason, they have not been used in this evaluation study.

\section{Results} \label{sec:results}
For the comparative analysis, the experimentation was run on a CPU i7-7700HQ 2.8GHz, 16GB RAM DDR4, on Windows 10 64bit. The maximum threshold allowed for the suppression technique required by the stakeholder is 0.5\%. The computational time limitations were set to 15 hours. Others metaheuristic parameters related to \textsc{KGen} and Random-Search Approach can be seen in Tab. \ref{tab:metaheuristic_parameter}.

\begin{table}[ht!]
    \centering
    \renewcommand{\arraystretch}{1.5}
    \caption{\textit{Metaheuristic parameters setup.}}
    \resizebox{.5\linewidth}{!}{
        \begin{tabular}{lll}
            \hline
            & KGEN & Random \\
            \hline
            maxEvaluations & 5000 & 5000 \\
            populationSize & 100 & 5000 \\
            crossoverRate & 0.9 & - \\
            mutationRate & 0.2 & - \\
            horizontalMutationRate & 0.4 & - \\
            \hline
        \end{tabular}
    }
    \label{tab:metaheuristic_parameter}
\end{table}


\subsection{RQ1: \textsc{KGen} Performance}


\begin{figure}[h!]
\centering
    \begin{subfigure}[h]{.7\textwidth}
    \includegraphics[width=\textwidth]{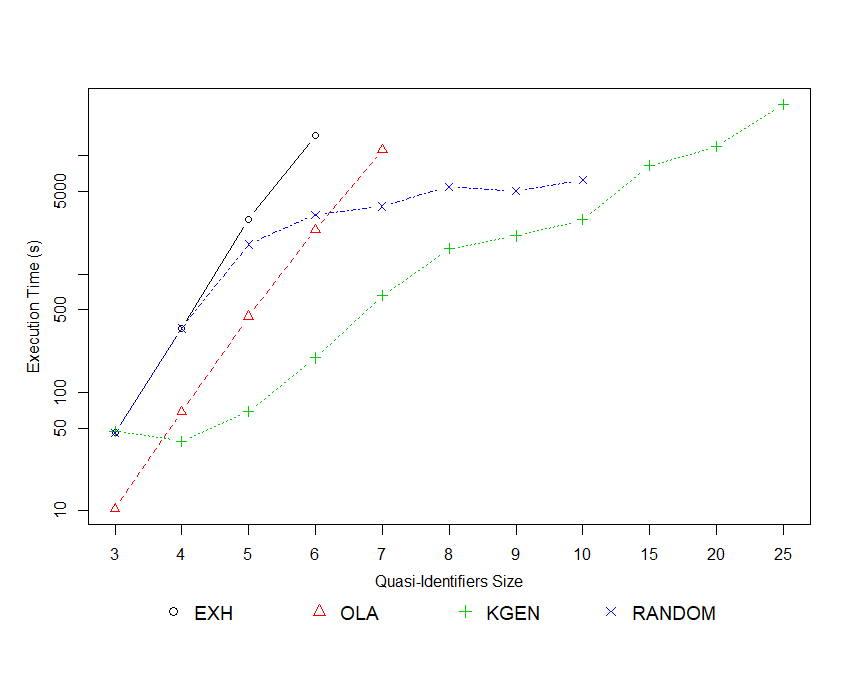}
    \caption{\footnotesize Execution time c2k dataset.}
    \label{fig:execution_time_ld}
\end{subfigure}
\begin{subfigure}[h]{.7\textwidth}
    \includegraphics[width=\textwidth]{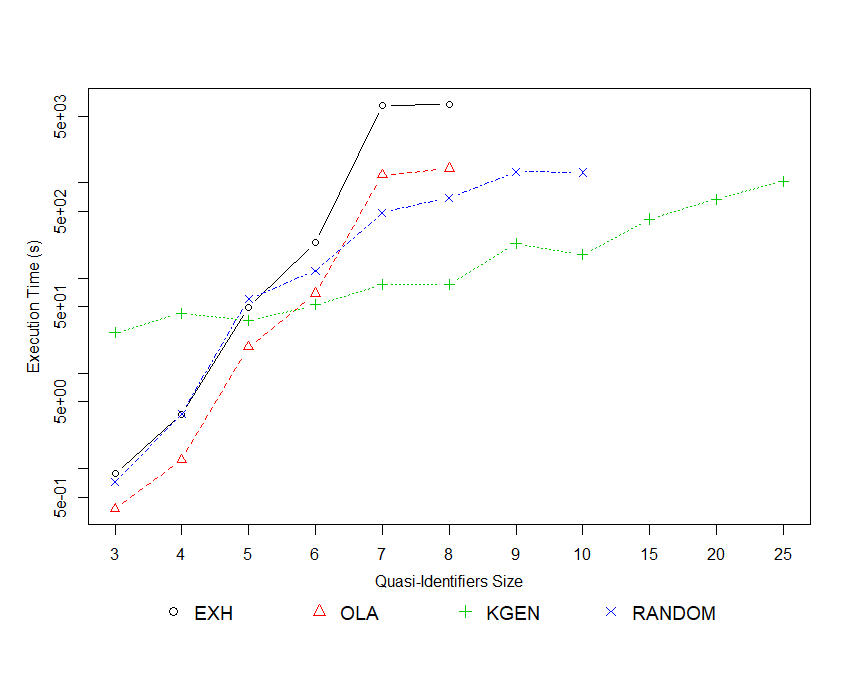}
    \caption{\footnotesize Execution time \textcolor{\editcolor}{F$^{2}$ dataset}.}
    \label{fig:execution_time_f2}
\end{subfigure}
\caption{\footnotesize Execution time evaluation results over the considered datasets.}
\label{fig:execution_time}
\end{figure}

Fig. \ref{fig:execution_time} plots execution times in a logarithmic scale. The exact approach can give results for a maximum of 6 QID for the c2k dataset and 10 QID for F$^{2}$ dataset while its computation halts or crashes with the increase of QIDs. Conversely, \textsc{KGen} and random-search provide results until to 25 QID for c2k and 15 QID for F$^{2}$.

\subsection{RQ2: \textsc{KGen} Accuracy}


\begin{figure}[h!]
\centering
    \begin{subfigure}[h]{.7\textwidth}
    \includegraphics[width=\textwidth]{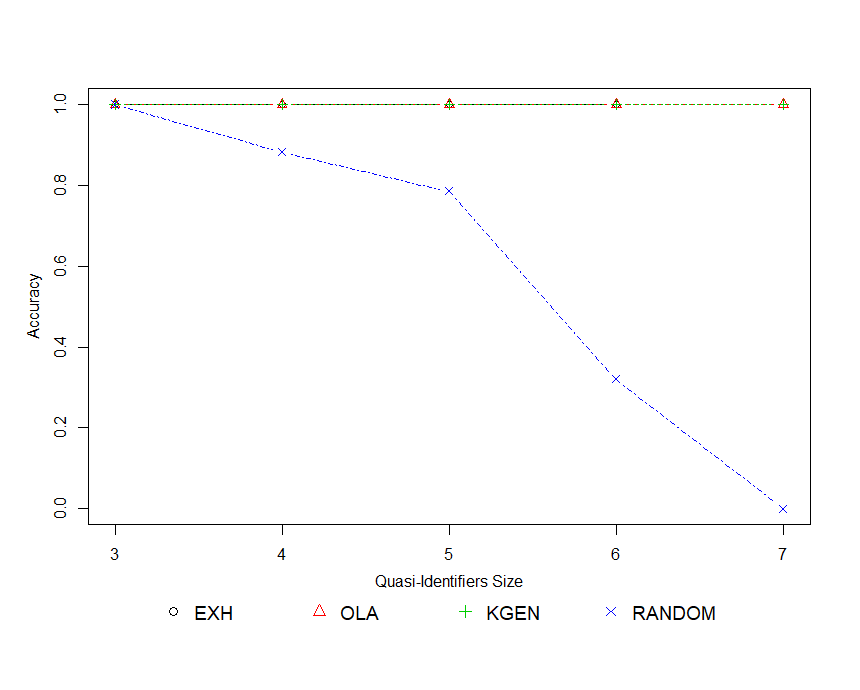}
    \caption{\footnotesize Accuracy on c2k dataset.}
    \label{fig:acc_ld}
\end{subfigure}
\begin{subfigure}[h]{.7\textwidth}
    \includegraphics[width=\textwidth]{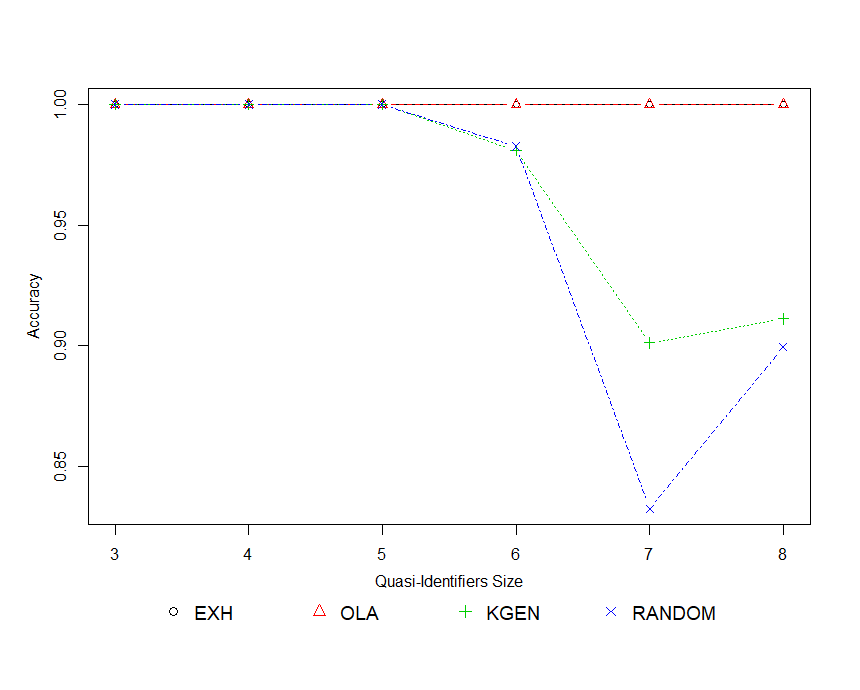}
    \caption{\footnotesize Accuracy on \textcolor{\editcolor}{F$^{2}$ dataset}.}
    \label{fig:acc_f2}
\end{subfigure}
\caption{Accuracy evaluation results over the considered datasets.}
\label{fig:acc}
\end{figure}



\begin{figure}[h!]
\centering
    \begin{subfigure}[h]{.7\textwidth}
    \includegraphics[width=\textwidth]{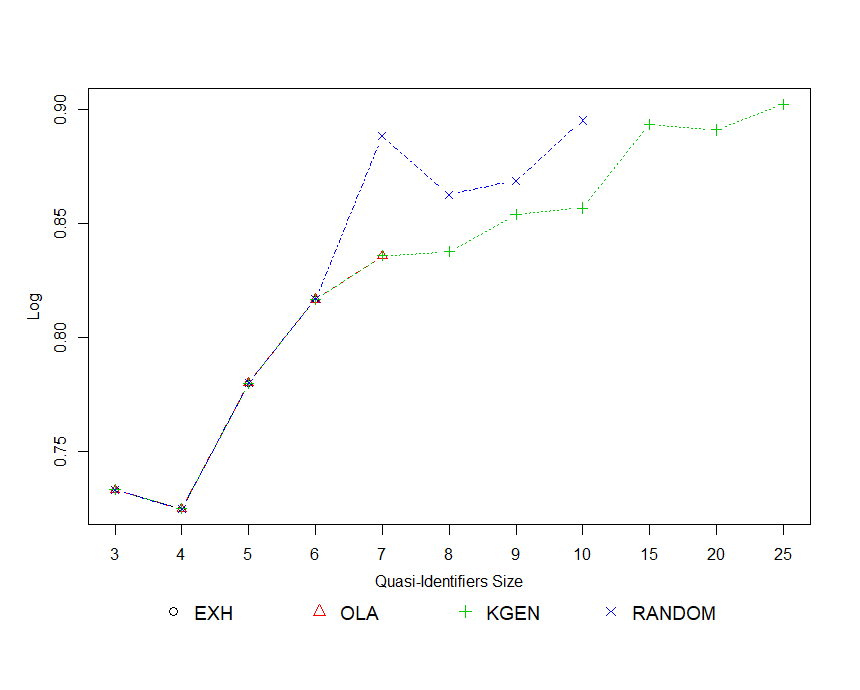}
    \caption{\footnotesize Solution quality, Level of generalization (LOG) of c2k dataset.}
    \label{fig:log_ld}
\end{subfigure}
\begin{subfigure}[h]{.7\textwidth}
    \includegraphics[width=\textwidth]{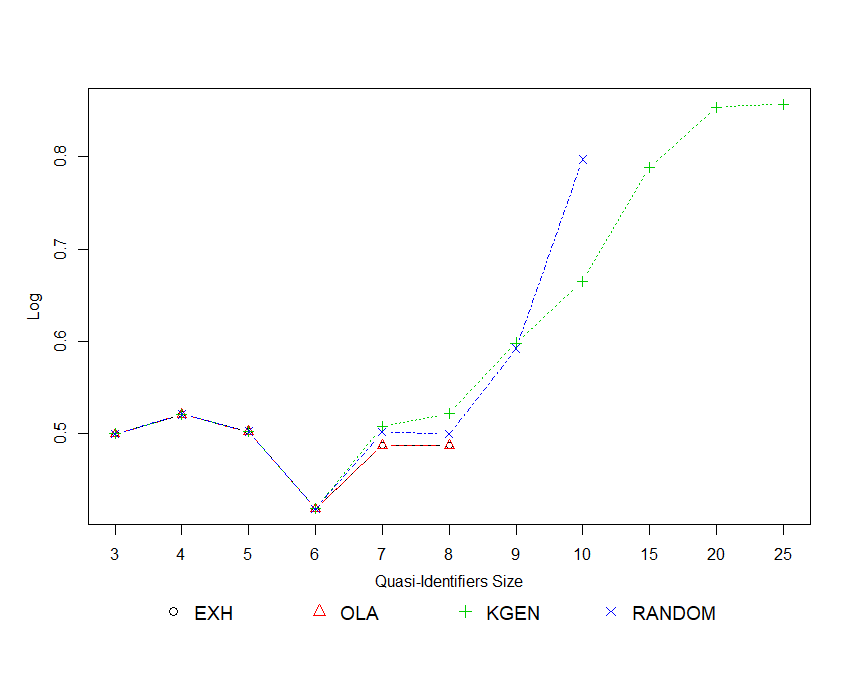}
    \caption{\footnotesize Solution quality, Level of generalization (LOG) of \textcolor{\editcolor}{F$^{2}$ dataset}.}
    \label{fig:log_f2}
\end{subfigure}
\caption{Level Of Generalization on the dataset anonymized.}
\label{fig:log}
\end{figure}

\begin{figure}[h!]
\centering
    \begin{subfigure}[h]{.7\textwidth}
    \includegraphics[width=\textwidth]{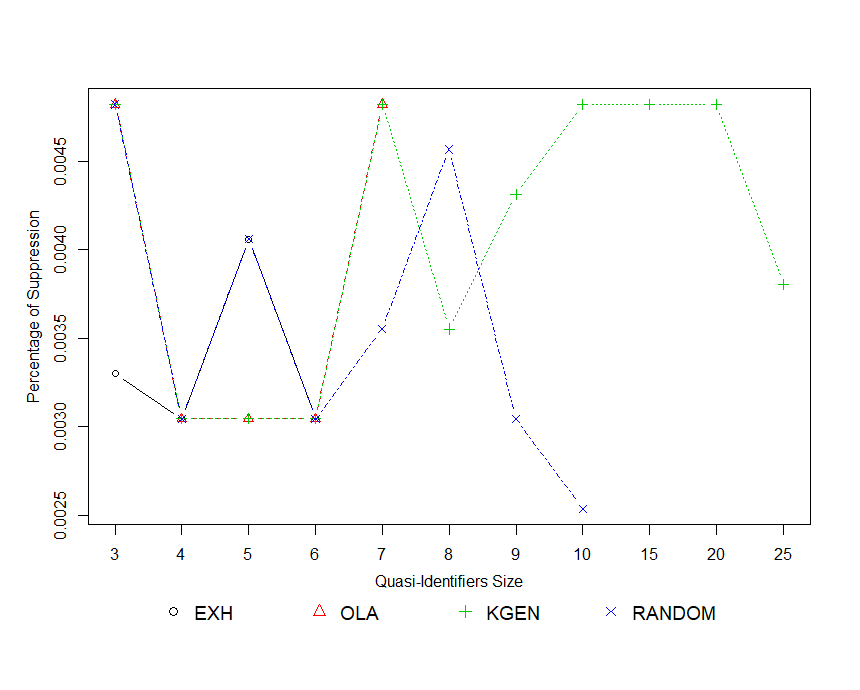}
    \caption{\footnotesize Level of suppression (LOS) of c2k dataset.}
    \label{fig:sup_ld}
\end{subfigure}
\begin{subfigure}[h]{.7\textwidth}
    \includegraphics[width=\textwidth]{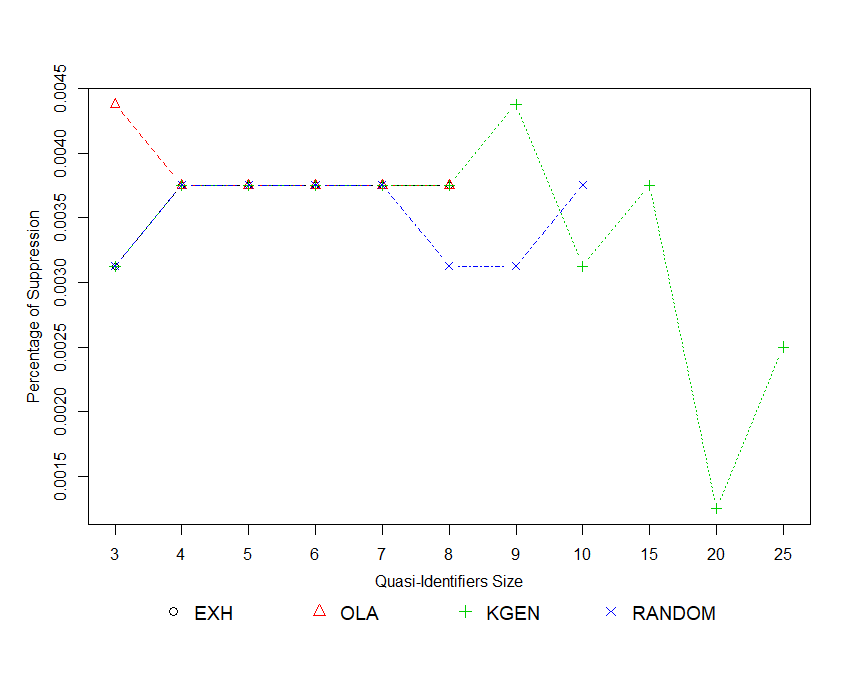}
    \caption{\footnotesize Level of suppression (LOS) of \textcolor{\editcolor}{F$^{2}$ dataset}.}
    \label{fig:sup_f2}
\end{subfigure}
\caption{\footnotesize Level Of Suppression on the dataset anonymized.}
\label{fig:sup}
\end{figure}

Fig. \ref{fig:acc} outlines results for accuracy. Given the limitation of the exact approaches to provide the optimal solution for several quasi-identifiers higher than 7 for the c2k dataset and 8 for the real dataset, the accuracy graph shows the accuracy level only up to 7 or 8 quasi-identifiers.

As the left-hand side of the figure shows, most approaches, including \textsc{KGen} offer accurate results with the apparent exception of the random approach, which, by definition, is bound to be non-accurate. On the right-hand side, the real data dataset results show that the accuracy decreases from the seventh quasi-identifiers.

\subsection{RQ3: \textsc{KGen} solution quality}

Fig. \ref{fig:log} and Fig. \ref{fig:sup} show the level of generalization and suppression of all approaches compared. In the scope of the plot, to evaluate the extent of goodness for approaches different than exact ones (i.e., \textsc{KGen} and random), it is sufficient to evaluate how 
\textcolor{\editcolor}{low are their curves.}

\textcolor{\editcolor}{Regarding the level of generalization, the \textsc{KGen} result, except with the F$^{2}$ dataset with 7/8 quasi-identifiers, is always equal or lower to the other approaches. Even when the other approaches cannot provide a solution, \textsc{KGen} provides better results than the Random approach.}

\textcolor{\editcolor}{The suppression criteria, instead, presents a different behaviour depending on the dataset used. With the F$^{2}$ dataset, the behaviour of 
\textsc{KGen} seems to be the same as the exact approaches, and the suppression value with a number of quasi-identifiers higher than 9 seems to decrease. The c2k dataset, instead, presents curves with unstable behaviour for all the approaches considered, making it more challenging to analyze. Nonetheless, the \textsc{KGen} behaviour is equal to the exact approaches. Considering that exact approaches provide the best results, it means that \textsc{KGen} provides the same good results as the exact approaches results.}

\section{Discussion} \label{sec:discussion}
As expected from our results on RQ1
the state-explosion problem \cite{clarke08} does not allow to have the exact solution in a reasonable time in all cases. More specifically, with more than 6 quasi-identifier attributes for the c2k dataset and 10 quasi-identifier attributes for the F$^2$ dataset, it is unfeasible to run exact approaches. Differently, with the usage of metaheuristics, we can provide solutions until 25 quasi-identifiers attributes and opportunistically continue if granted with the appropriate computational means. Clearly, from that point onwards, also for metaheuristic approaches is difficult to provide a solution. One factor strongly related to the increasing of the execution time on metaheuristics pertains to the maximum number of evaluations, based on metaheuristic configuration (e.g., see Tab. \ref{tab:metaheuristic_parameter}) since the number of nodes evaluated is directly related to these configurations. Consequently, to decrease the execution time, operators and data processing agents can fine-tune the maxEvaluation parameter of \textsc{KGen} (or even the random approach) opportunistically and as needed. Another important aspect is that the slope of execution-time curves for the random approach is lower than \textsc{KGen} at the increase of QID. This is because a single evaluation run in \textsc{KGen} analyzes more than one single node, given that the crossover operator continuously generates new nodes. This limitation can be the object of future study by researchers and practitioners interested to address its impact.

Moreover, concerning  RQ2, the accuracy level shows how \textsc{KGen} provides solutions that are identical $\pm$.9\% to the optimal approach. It means that \textsc{KGen} can (a) converge using its genetics operators to the optimal solution with small instances and (b) to be very close to the optimum as the number of instances increases.
Conversely, the random approach initially provides a good level of accuracy due to the number of evaluations concerning the size of the problem. For example, if a lattice contains 300 nodes, with 5000 evaluations (setting of the random approach described in Tab. \ref{tab:metaheuristic_parameter}), the random approach analyzes all nodes in the lattice, providing a high level of accuracy. However, the opposite is true exponentially with the increase of lattice nodes.

Focusing on RQ3, we can observe the level of generalization and suppression of \textsc{KGen} as being very close to the level of generalization of optimal approaches. This is a good indicator of the power of our research solution. Most notably, our approach (just as the random one) can provide solutions that can deal with higher numbers of quasi-identifier attributes, a feature where most optimal approaches fail. Unlike the random approach, however, \textsc{KGen} provides excellent results in terms of generalization level, considering the suppression applied. If the random approach seems to have better results on large instances, considering only the generalization level, we can see how this is due to the high level of suppression applied by the random approach itself. Looking at both metrics, we can easily understand how \textsc{KGen} has the best results.

From the results of the three sub-research questions, we can assume that \textsc{KGen} performs well in real case contexts. 
Moreover, given the dataset provided by the Taskforce West-Brabant-Zeeland, we can anonymize their dataset with a good level of anonymization, having the same results of exact approaches.

Unlike heuristic approaches, meta-heuristic approaches can also perform well in a context where the dataset size, in terms of the number of quasi-identifiers, is more extensive. \textcolor{\editcolor}{Hence, a stakeholder can use \textsc{KGen} in large contexts scenarios.} 
\textcolor{\editcolor}{Nonetheless, to ensure the applicability in a general context, the approach needs to be validated with more datasets.} 

\textcolor{\editcolor}{Lastly, after providing the anonymized dataset, KGen provides also metadata regarding the information loss for each dataset attribute. Hence, the final user can estimate the damage entity by mean of information loss of each attribute.}
\section{Limitations and Threats to Validity} \label{sec:ttv}

\begin{table}
\footnotesize
    \renewcommand{\arraystretch}{1.5}
    \caption{\footnotesize Lattice reduction process with suppression criteria.}
    \begin{subtable}{\linewidth}
        \centering
        \begin{tabular}{ccc}
            \hline
            \textbf{Age} & \textbf{Postcode} & \textbf{Gender}\\
            \hline
            24 & 80015 & F \\
            28 & 80019 & M \\
            42 & 85073 & F \\
            \hline
        \end{tabular}
        \caption{\footnotesize Example of a dataset not k-anonymized.}
        \label{tab:ttv_dataset}
    \end{subtable}
    
    \begin{subfigure}{\linewidth}
        \begin{subtable}{.33\linewidth}
          \centering
            \begin{tabular}{c}
                \hline
                \textbf{Age} \\
                \hline
                20 - 29 \\
                20 - 29 \\
                \sout{40 - 49} \\
                \hline
            \end{tabular}
            \caption{Age: LOG 2.}
            \label{tab:ttv_age}
        \end{subtable}%
        \begin{subtable}{.33\linewidth}
          \centering
            \begin{tabular}{c}
                \hline
                \textbf{Postcode} \\
                \hline
                8001* \\
                8001* \\
                \sout{8507*} \\
                \hline
            \end{tabular}
            \caption{PC: LOG 1.}
            \label{tab:ttv_pc}
        \end{subtable} 
        \begin{subtable}{.33\linewidth}
          \centering
            \begin{tabular}{c}
                \hline
                \textbf{Gender} \\
                \hline
                F \\
                \sout{M} \\
                F \\
                \hline
            \end{tabular}
            \caption{Gender: LOG 0}
            \label{tab:ttv_gender}
        \end{subtable}
        \caption{\footnotesize Datasets k-anonymized, applying the suppression criteria (with a max level of suppression of 35\% of the entire dataset). The final dataset contains only the first row.}
        \label{tab:ttv_preprocessing_with_suppression}
    \end{subfigure}
    

    \label{tab:ttv_lattice_preprocessing}
    
\end{table}

This section outlines the major limitation we perceive in our work, which reflects one of the optimizations that \textsc{KGen} features in its processing and algorithms.
As outlined in section \ref{subsec:lattice_preprocessing}, \textsc{KGen} features a lattice size reduction technique that limits the approach applicability in specific cases. Nevertheless, the technique is essential since it can work on a smaller search space than the original one, whose size could be untractable without major software-defined infrastructure requirements. However, the described technique introduces a vulnerability when, during the anonymization process, also the suppression technique is introduced.
Preprocessing without suppression ensures that all lattice nodes except for the new minimum node found in the process are not k-anonymized. With the suppression active, instead, this does not hold.
Let us take into account the example in Tab. \ref{tab:ttv_lattice_preprocessing}. If we apply the suppression criteria (with a maximum level of suppression set, by default, of 35\%) on each dataset, in order that all datasets are k-anonymized, we have the suppression of the last row on the first and second dataset and the suppression of the second row in the last dataset (Tab. \ref{tab:ttv_age} - \ref{tab:ttv_pc} - \ref{tab:ttv_gender}). At this point, removing the second and third-row from the dataset, the remaining dataset is composed by only the first row, with a final level of generalization of $<$2, 1, 0$>$.
Nevertheless, this dataset is k-anonymized also without any generalization ($<$0, 0, 0$>$). 
By applying the suppression criteria, it is possible to have one k-anonymized node with a level of generalization less than the minimum level of generalization provided by the preprocessing.
We are aware of this limitation and plan to address it in future developments and iterations over this work.
\section{Conclusion and future work}\label{sec:conclusion}
With the quickly increasing amount of digital data, there emerges a growing need to provide support for fast and scalable data-processing capable of offering anonymization guarantees.  
In this paper, we introduce \textsc{KGen}, a scalable approach to data-intensive k-anonymization featuring genetic algorithms. 

The \textsc{KGen} approach focuses on the assessment of the balance between two critical, and opposite data quality attributes functional to data-processing, namely, data \emph{privacy} versus \textcolor{\editcolor}{\emph{usefulness of data}}. As aforementioned, \textsc{KGen} exploits genetic algorithms that allow organically increasing the level of privacy of the data while safeguarding that the data evidence, which is still \emph{usable} e.g., in terms of financial evidence and audit trails part of governmental data-intensive processing.

\textsc{KGen} is 
a practical, scalable, data-intensive approach that can effectively anonymize datasets embracing the well-accepted k-anonymization measure. The approach is supported by a prototype coded in Java and tested through various experiments using benchmarks and real-life industrial datasets. 

Initial results look very promising. We have shown empirically that the behavior of \textsc{KGen} level of generalization metric performance equally well as other optimization approaches, while \textsc{KGen} -in contrast to other approaches- can deal with a large number of quasi-identifiers, and thus Big datasets. 

Future work will focus on building a more robust and user-friendly interface on top of the current prototype and more personalized privacy measures. Besides, we intend to work on a dynamic version of \textsc{KGen}, D-\textsc{KGen}, that can deal with streaming data that dynamically add/remove/alter the dataset on-the-fly and just-in-time, breaking the ``closed-world assumption'' underpinning most of the existing approaches.

\bibliographystyle{elsarticle-num} 
\bibliography{main.bib}

%

\end{document}